\documentclass[preprint,proceedings]{rmaa}

\suppressfulladdresses 

\SetYear{2006}
\SetConfTitle{First Light Science with the GTC}

\title{The ``Super-Halo" of M31 and M33} 

\author{
  Ata Sarajedini,\altaffilmark{1}
  }

\altaffiltext{1}{University of Florida, Department of Astronomy, 211 Bryant Space 
Science Center, Gainesville, FL 32611-2055 
(ata@astro.ufl.edu)}

\shortauthor{Sarajedini}
\shorttitle{M31 and M33 Super-Halo}

\listofauthors{Ata Sarajedini}
\indexauthor{Sarajedini, A.}

\abstract{Two recent observations regarding the halo of M33 seem to contradict each other.
First, the star clusters in the halo of M33 exhibit an age range of 5 to 7 Gyr suggesting a
formation scenario that involves the chaotic fragmentation and accretion of dwarf
satellites. In contrast, deep photometric searches for the resultant tidal tails and stellar 
streams in the vicinity of M33 have turned up nothing significant. In this contribution,
we have tried to reconcile these apparently disparate observations.
 We suggest that M33 is situated within
a `superhalo' which contains many other dwarf spheroidal and dwarf irregular galaxies
that are satellites of M31. In such a scenario, the tidal field of M31 could have disrupted
and/or diluted the leftover tails and streams leaving little to be detected in the present
day.}

\addkeyword{galaxies: dwarf}
\addkeyword{galaxies: spiral}
\addkeyword{galaxies: structure}
\addkeyword{galaxies: formation}

\begin{document}
\maketitle

\section{Introduction}
The Local Group of galaxies is an excellent laboratory within which to conduct detailed
studies of how galaxies form and evolve. One reason for this is that the galaxies are 
all close enough to make in-depth studies of their stellar populations feasible.
Another reason is that the Local Group contains examples of all major galaxy 
morphology classes - including spirals, ellipticals, dwarf irregulars, and dwarf spheroidals.
There is enough diversity in the Local Group so that studies of the detailed
properties of its galaxies hold great promise in unlocking the secrets of star and galaxy
formation in a variety of different environments. In addition, the Local Group is
also a window into the general process of structure formation in the early Universe
and the relevance of cold dark matter models in describing this process (e.g. Font et al. 2006,
and references therein).

In this contribution, we focus on two spiral galaxies of the Local Group - M31
and M33. At a distance of $\sim$800 kpc, the M31 system includes over a dozen dwarf galaxies,
including M33, which is a member of the class of {\it dwarf spirals}. It is 
approximately one magnitude brighter than the Large Magellanic Cloud. Studies
have shown that M33 possesses kinematically distinct disk and halo populations 
(Schommer et al. 1991; Chandar et al. 2002).
The halo of M33 is particularly intriguing because its member star clusters exhibit an  
age range of between 5 and 7 Gyr (Sarajedini et al. 1998, 2000; Chandar et al. 2002).  
While the spatial extent of the M33 disk and 
halo is still largely unknown, wide-field photometric surveys of M33 do not show any traces of
tidal tails or streams in its halo (Fig. 3 of McConnachie et al. 2004), unlike the Milky
Way (Ibata, Gilmore, \& Irwin 1994; Majewski et al. 2006) and M31 (Ferguson et al. 2002;
Guhathakurta et al. 2006), which exhibit extensive substructure in their halos. It
should be noted however that in their kinematical study of a small region along the
major axis of M33, McConnachie et al. (2006) detect a minor velocity component that 
is neither coincident with the disk nor the halo, and which they suggest could be a tidal
tail or stream. The significant age range of the M33 halo clusters (5-7 Gyr) suggests a
formation scenario for the M33 halo which involves the fragmentation and accretion of 
subgalaxy fragments. This is at odds with the finding that the remnants of such objects 
are largely missing today in the M33 halo. This apparent contradiction can be better
understood if we examine M33 as a member of the M31/M33 superhalo.

\begin{figure}[t!]
\includegraphics[width=\columnwidth]{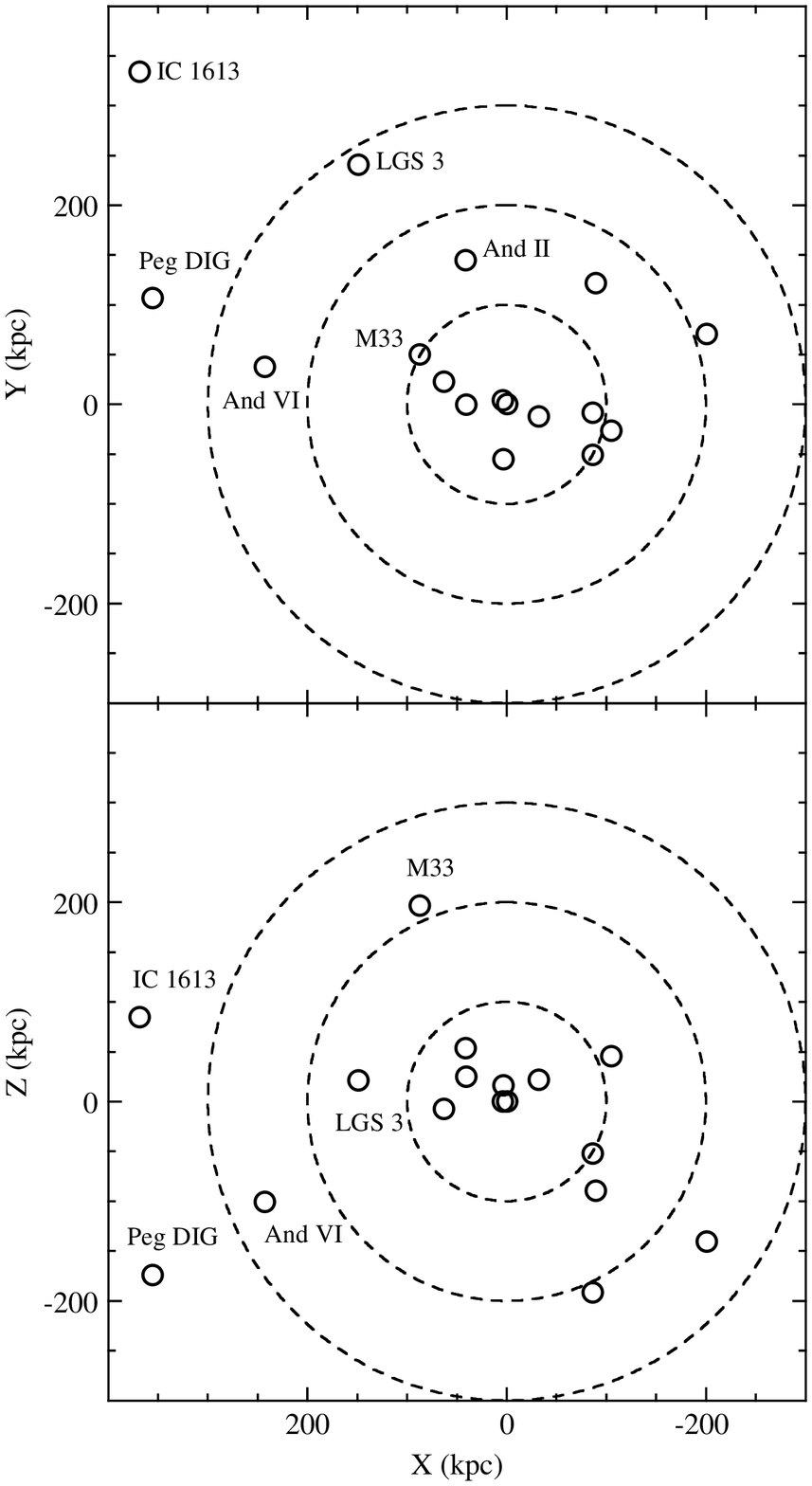}
\caption{The distribution of M31 dwarf galaxies and M33 in the X-Y-Z plane
defined with M31 at its center using data in Table 1 of 
Koch \& Grebel (2006). The concentric circles represent 100 kpc radii.}
\label{fig:psfrag}
\end{figure}

\section{M31 and M33 Super-Halo}

The M31 system of dwarf galaxies includes such notable members as M32, NGC 205,
NGC 185, NGC 147, and a number of dwarf spheroidal and dwarf irregular galaxies.
It also includes the dwarf spiral galaxy M33. Figure 1 shows the X-Y-Z locations of
galaxies in the superhalo of M31 centered on M31. These data are taken from Table 1 of 
Koch \& Grebel (2006). The dashed circles represent radii at 100 kpc intervals. Figure 1
shows that M33 is relatively isolated at a distance of $\sim$220 kpc from M31 in 
3-D space. 

Figure 2 shows the radial velocities of galaxies in the M31 superhalo taken from the
NASA Extragalactic Database as a function of radial distance in the disk of M31. 
These are the same galaxies that are plotted in Fig. 1. Figure 2 illustrates the fact
that the members of the superhalo are not rotating so that kinematically, they
represent a `hot' population. Note that the point located at the origin is M31.

We now turn to a discussion of Fig. 3 which plots the metallicities of various superhalo
constituents as a function of their distance from M31. The three open circles between 
10 and 100 kpc are the fields observed by Kalirai et al. (2006) from their Table 3. The
inner point is bulge-dominated, the outer point is halo dominated, while the middle
point is some mix of bulge and halo.
These fields are located to the southeast of M31 in the
direction of M33 on the sky. The open circle at 1.6 kpc 
represents the putative M31 bulge from the work of Sarajedini \& Jablonka (2005).
The dashed line is the least squares fit to the four open circles. The filled circles are
the dwarf galaxies in Table 1 of Koch \& Grebel (2006) supplemented by the metallicities
given in Table 1 of Grebel, Gallagher, \& Harbeck (2003). The metallicity dispersions
in Table 1 of Grebel et al. (2003) have been divided in half to simulate 1-$\sigma$ 
errors. We also include the location of And IX published in the work of
Harbeck et al. (2005). These metallicities are on an internally consistent scale and have
been estimated via photometric and spectroscopic methods (Grebel et al. 2003).
The squares represent three globular clusters in M31 -
B327 and G1 (Koch \& Grebel 2006, open squares) and the furthest known
globular cluster from the center of M31 discovered by Martin et al. (2006, filled
square). Lastly, the rectangular region is the location and metallicity range of
halo clusters in M33 based on the work of Sarajedini et al. (1998; 2000).

\section{Discussion}

It is interesting to note that the dashed line in Fig. 3 fitted to the open circles
in M31 also intersects the low metallicity regime of the {\it M33 halo clusters}. The globular
cluster G1 and the most distant M31 globular are also consistent with this line. The
globular cluster B327 and the dwarf elliptical galaxy M32 are far from the dashed
line, which could be because their current positions (close to the center of M31) 
are not representative of their apogalactic locations (i.e. their formation locations). 
With the possible exception of And IX, the dwarf spheroidals and
irregulars also cluster around this line but whether they actually follow the dashed line
is an open question. 

Focusing now on the location of M33 in Fig. 3, one could argue that
because the metal-poor tail of the halo globular clusters in M33 is consistent with the
``bulge" metallicity gradient of M31, that the former is actually a constituent of the latter.
On the other hand, one could also claim that because the entire range of halo metallicities 
in M33 is within the range of the M31 dwarf spheroidals and irregulars, then
the halo of M33 is a constituent of the M31 halo not its ``bulge." In either case, Fig. 3
suggests that M31 and M33 reside in a superhalo that also contains at least 15
other galaxies of various types.

This concept of a superhalo containing M31 and M33 along with other galaxies can
potentially explain the apparent contradiction presented in the Introduction. On the 
one hand, there is a
significant age range among the halo star clusters in M33 suggesting that a process
akin to fragmentation and accretion formed the M33 halo. If so, then we should see
traces of tidal tails and streams that represent the `leftovers' of this process. On the
other hand,
surveys to search for these signatures (at the same surface brightness levels as those
in M31) have revealed nothing significant (e.g. McConnachie et al. 2004). This 
indicates that the streams could be
below the surface brightness detection limits of the surveys, which can happen
if the tidal field of the superhalo has disrupted and/or diluted the streams. In 
addition, this suggests that streams located in another part of the superhalo could
have originated as dwarf galaxies in the vicinity of M33. The properties of such
a stream such as metallicity and age would be more similar to those of M33 than M31.
All of these speculations depend on the strength of the M31 tidal field at the
location of M33, which is a highly uncertain quantity. Not only are the relative
masses of M31 and M33 uncertain, but the relative positions of M31 and M33 in 
space when the bulk of the tidal stream destruction was occurring is also uncertain.

\begin{figure}
\includegraphics[width=\columnwidth]{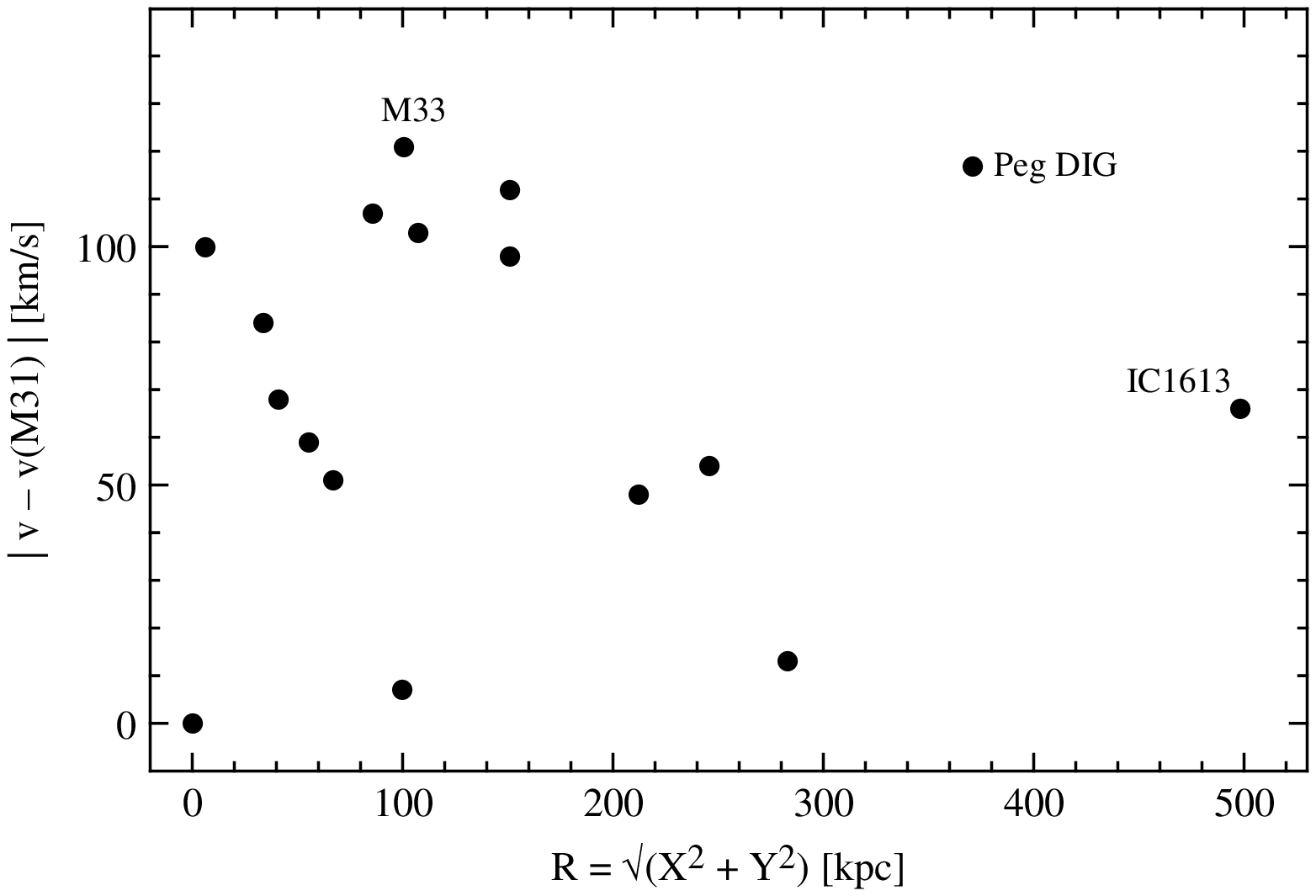}
\caption{The distribution of radial velocities relative to M31 for the M31
dwarf galaxies and M33. The positional data are taken from Table 1 of 
Koch \& Grebel (2006) while the velocities are from the NASA Extragalactic Database.}
\label{fig:psfrag}
\end{figure}

\begin{figure}[t!]
\centering
\includegraphics[width=\textwidth,height=110mm,width=160mm]{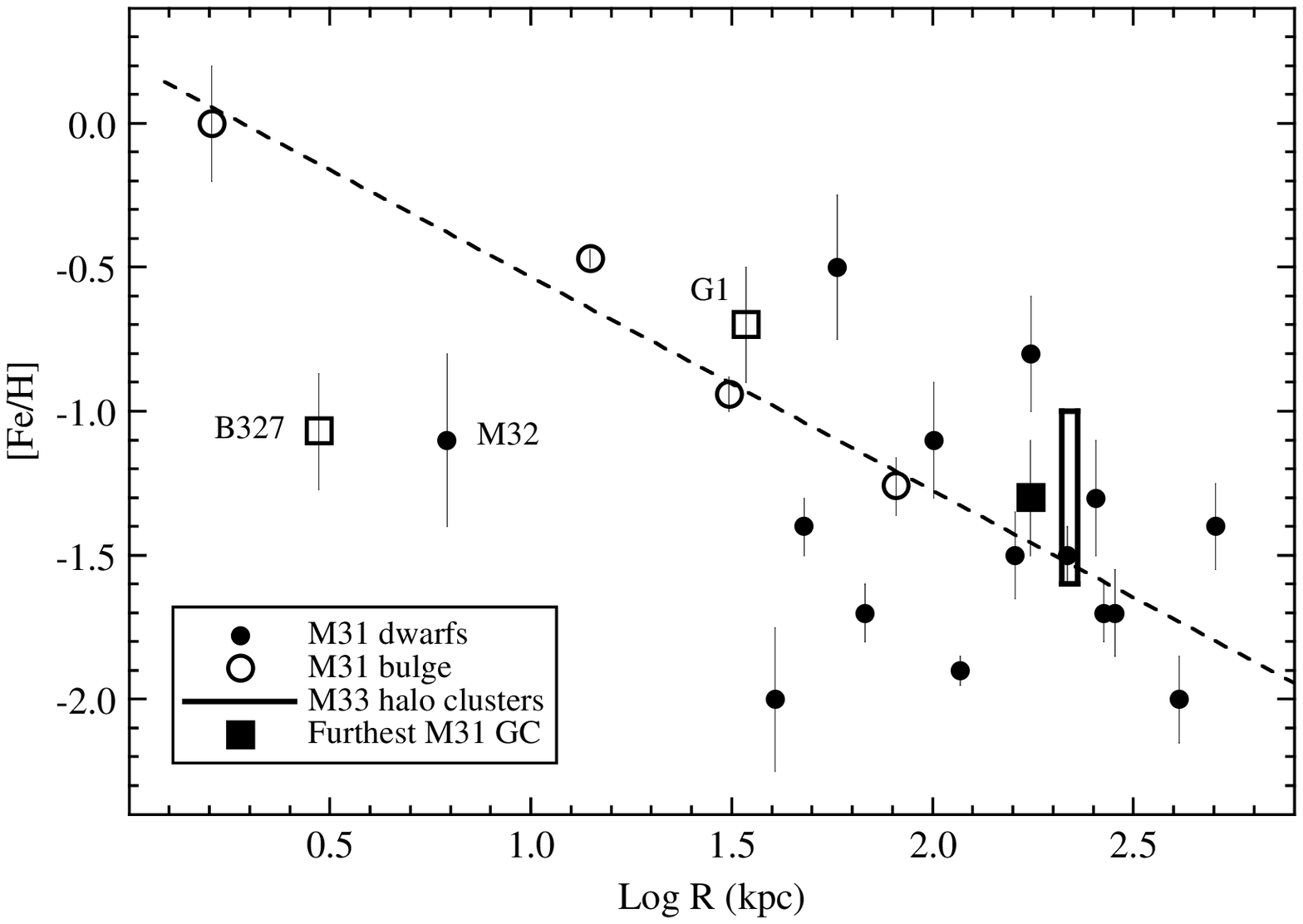}
\onecolumn
\caption{The radial metallicity gradient in the super-halo of M31. The open
circles between 10 and 100 kpc are taken from Kalirai et al. (2006)
while the one representing the putative bulge population is from Sarajedini \& Jablonka
(2005).  The dashed line is the least squares fit to these data. The filled circles are
the dwarf galaxies in Table 1 of Koch \& Grebel (2006) supplemented by the metallicities
given in Table 1 of Grebel, Gallagher, \& Harbeck (2003). We also include the location 
of And IX (Harbeck et al. 2005). The squares are three globular clusters in M31 
(Koch \& Grebel 2006, open squares; Martin et al. (2006, filled
square). The rectangular region is the location and metallicity range of
halo clusters in M33 based on the work of Sarajedini et al. (1998; 2000).}
\twocolumn
\end{figure}

\section{Summary and Conclusions}

In this short contribution, we have tried to reconcile the significant age range
among M33 halo clusters which argues in favor of halo formation via fragmentation/accretion 
of dwarf satellites and the lack of detectable tidal streams  in the vicinity 
of M33 that would be leftover from such a process. We show that M33 is situated within
a `superhalo' which contains many other dwarf spheroidal and dwarf irregular galaxies
that are satellites of M31. We suggest that, because
M33 is within the superhalo of M31, the tidal field of the latter has successfully disrupted
and/or diluted the leftover tails and streams thereby rendering them undetectable
by imaging surveys that have found similar structures near M31.

\section{Acknowledgements}

The author would like to acknowledge useful comments from Mike Barker, Daniel
Harbeck, Aaron Grocholski, Rupali Chandar, Jason Kalirai, and Alan McConnachie 
on an earlier version of this manuscript. This work was supported by NSF CAREER grant 
AST-0094048.


\begin{thebibliography}
\bibitem{} Chandar, R., Bianchi, L., Ford, H. C. \& Sarajedini, A. 2002, ApJ, 564, 712
\bibitem{} Ferguson, A. 2002, AJ, 124, 1452
\bibitem{} Font, A. S., Johnston, K. V., Bullock, J. S., Robertson, B. E. 2006, ApJ, 646, 886
\bibitem{} Grebel, E. K., Gallagher, J. S. III, \& Harbeck, D. 2003, AJ, 125, 1926
\bibitem{} Guhathakurta, P. et al. 2006, AJ, 131, 2497
\bibitem{} Harbeck, D., Gallagher, J. S., Grebel, E. K., Koch, A., \& Zucker, D. B.,
2005, ApJ, 623,159
\bibitem{} Ibata, R. A., Gilmore, G., \& Irwin, M. J. 1994, Nature, 370, 194
\bibitem{} Kalirai, J. S. et al. 2006, ApJ, 648, 389
\bibitem{} Koch, A., \& Grebel, E. K. 2006, 131, 1405
\bibitem{} Majewski, S. R., Law, D. R., Polak, A. A., \& Patterson, R. J. 2006, ApJ,
637, L25
\bibitem{} Martin, N. F. et al. 2006, MNRAS, 371, 1983
\bibitem{} McConnachie, A. W. et al. 2004, MNRAS, 350, 243
\bibitem{} McConnachie, A. W. et al. 2006, ApJ, 647 L25
\bibitem{} Sarajedini, A., \& Jablonka, P. 2005, AJ, 130, 1627
\bibitem{} Sarajedini, A., Geisler, D., Harding, P., \& Schommer, R. 1998, ApJ, 508, L37
\bibitem{} Sarajedini, A., Geisler, D., Schommer, R.  \& Harding, P. 2000, AJ, 120, 2437
\bibitem{} Schommer, R. A., Christian, C. A., Caldwell, N., Bothun, G. D., \&
Huchra, J. 1991, AJ, 101, 873
\end{thebibliography}
\end{document}